%% file: main.tex
\documentclass[conference]{IEEEtran}
\IEEEoverridecommandlockouts
\usepackage{cite}
\usepackage{amsmath,amssymb,amsfonts}
\usepackage{graphicx}
\usepackage{textcomp}
\usepackage{xcolor}
\usepackage{multirow}
\def\BibTeX{{\rm B\kern-.05em{\sc i\kern-.025em b}\kern-.08em
    T\kern-.1667em\lower.7ex\hbox{E}\kern-.125emX}}
\newtheorem{myDef}{Definition}
\usepackage{listings}
\usepackage{fancybox}

\usepackage{algorithm}
\usepackage{algpseudocode}
\algblock{Input}{EndInput}
\algnotext{EndInput}
\algblock{Output}{EndOutput}
\algnotext{EndOutput}
\newcommand{\Desc}[2]{\State \makebox[2em][l]{#1}#2}

\usepackage{subfigure}
\usepackage{graphicx}
\usepackage{float}

\begin{document}

\title{Comprehensive Integration of API Usage Patterns}
\author{
\IEEEauthorblockN{
Qi Shen,
Shijun Wu,
Yanzhen Zou\IEEEauthorrefmark{1}\thanks{\IEEEauthorrefmark{1}Corresponding author.}, and
Bing Xie
}

\IEEEauthorblockA{Key Laboratory of High Confidence Software 
Technologies (Peking University), Ministry of Education\\}
\IEEEauthorblockA{Institute of Software, EECS, Peking University, Beijing, China}
\{shenqi16,wushijun,zouyz,xiebing\}@pku.edu.cn
}
 
\maketitle

\begin{abstract}
Nowadays, developers often reuse existing APIs to implement their programming tasks.
A lot of API usage patterns are mined to help developers learn API usage rules. 
However, there are still many missing variables to be synthesized when developers integrate the patterns into their programming context. 
To deal with this issue, we propose a comprehensive approach to integrate API usage patterns in this paper. 
We first perform an empirical study by analyzing how API usage patterns are integrated in real-world projects. We find the expressions for variable synthesis is often non-trivial and can be divided into 5 syntax types. Based on the observation, we promote an approach to help developers interactively complete API usage patterns. Compared to the existing code completion techniques, our approach can recommend infrequent expressions accompanied with their real-world usage examples according to the user intent. The evaluation shows that our approach could assist users to integrate APIs more efficiently and complete the programming tasks faster than existing works.
\end{abstract}


\input{sections/1_intro}
\input{sections/2_related}
\input{sections/3_analysis}
\input{sections/4_approach}
\input{sections/5_evaluation}
\input{sections/6_conclusion}

\input{sections/7_acknowledge}

\bibliographystyle{IEEEtran}
\bibliography{main.bib}

\end{document}

%% file: sections/1_intro.tex
\section{Introduction}
\label{section-intro}
Learning how to reuse existing APIs is a daily activity for developers.
Given an interested API method, developers often resort to online resources (\textit{e.g.}, Q\&A forums, blogs, tutorials) to search concrete API usage examples\cite{zhang2018code}.
Furthermore, many works\cite{raghothaman2016swim, wang2013mining, niu2017api} focus on mining abstract API usage patterns from large code corpus.
An API usage pattern is a code fragment mined from many concrete API usage examples, which documents that some API methods are frequently invoked in sequence to implement certain functionality.
Although API usage patterns are good starting points to learn APIs, it can be tedious for developers to integrate the APIs into their local programming context\cite{nguyen2010graph, zhang2019analyzing}. 
Existing studies show that a big barrier during the process is that many online code fragments are incomprehensive\cite{wu2019developers, treude2017understanding}.

\begin{figure}[htb]
	\centering
	\includegraphics[width=0.45\textwidth]{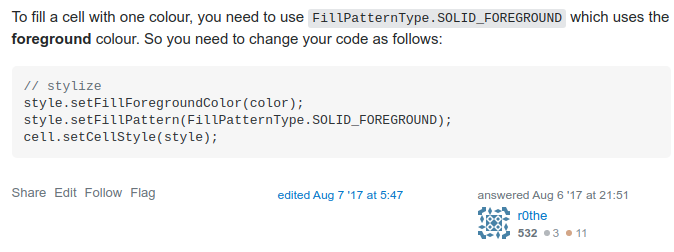}
	\caption{Incomplete API invocation sequence from online user forum}
	\label{fig1-1}
\end{figure}

Figure \ref{fig1-1} displays a code snippet from a Stack Overflow post\footnote{https://stackoverflow.com/questions/45536432}.
The code snippet displays the frequent API usage for setting the foreground color of an Excel cell with a third-party library \texttt{apache-poi}.
Although the snippet lists all important API methods to implement the functionality, it cannot be compiled or executed.
To complete the snippet, three well-typed expressions have to be manually synthesized for the \texttt{style}, \texttt{color}, and \texttt{cell} variables separately.

We summarize two challenges in the follow-up programming. 
The first one is variable understanding. 
Developers need to configure the APIs with correct local variables (\textit{i.e.}, receivers and parameters). 
However, the meaning of the missing variables is often not well explained in the code fragment. 
Each method invocation in Figure \ref{fig1-1} requires a receiver and a parameter. 
To specify the cell color in this task, users need to configure the parameters with type \texttt{short} in the second and third method calls. 
Yet the relation between a \texttt{short} variable and a user-specified color is quite subtle. 
The second challenge is variable synthesis.
After understanding the meaning of the variables to configure, developers need to synthesize a well-typed expression for each variable. 
The synthesized expression can be a constant, a constructor, or a method chain. 
For example, a correct expression to specify the red color in our example sequence is \texttt{IndexedColors.RED.getIndex()}. 
Although code completion is a common feature in modern IDEs (Integrated Development Environments), such completion typically considers one step of computation\cite{gvero2013complete}.
Furthermore, it is difficult for developers to judge the correctness of the recommended expressions when they are not familiar with the reused library.

\begin{figure}[htb]
\subfigure[User selects an API usage pattern]{
\begin{minipage}{8cm}
\centering
\includegraphics[width=\textwidth]{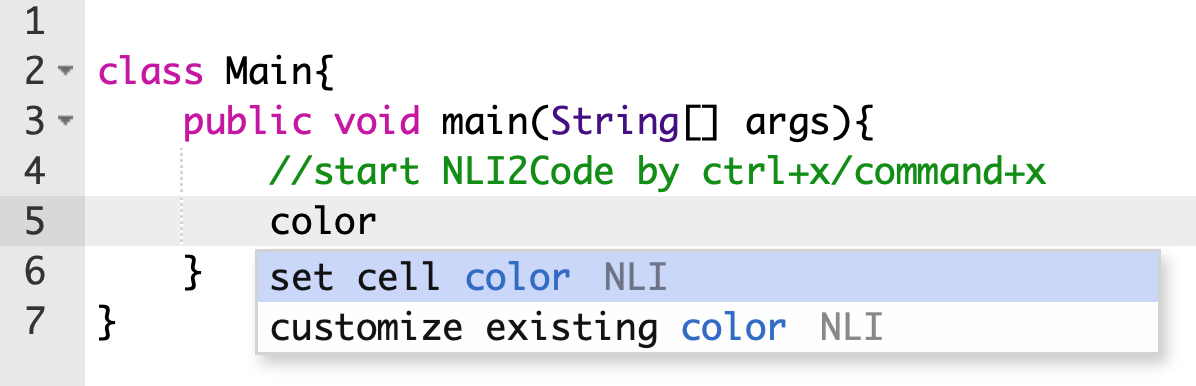}
\vspace{0.2cm}
\end{minipage}
}
\subfigure[\textsc{CodA} summarizes four parameters to configure, with recommended expressions]{
\begin{minipage}{8cm}
\centering
\vspace{0.2cm}
\includegraphics[width=\textwidth]{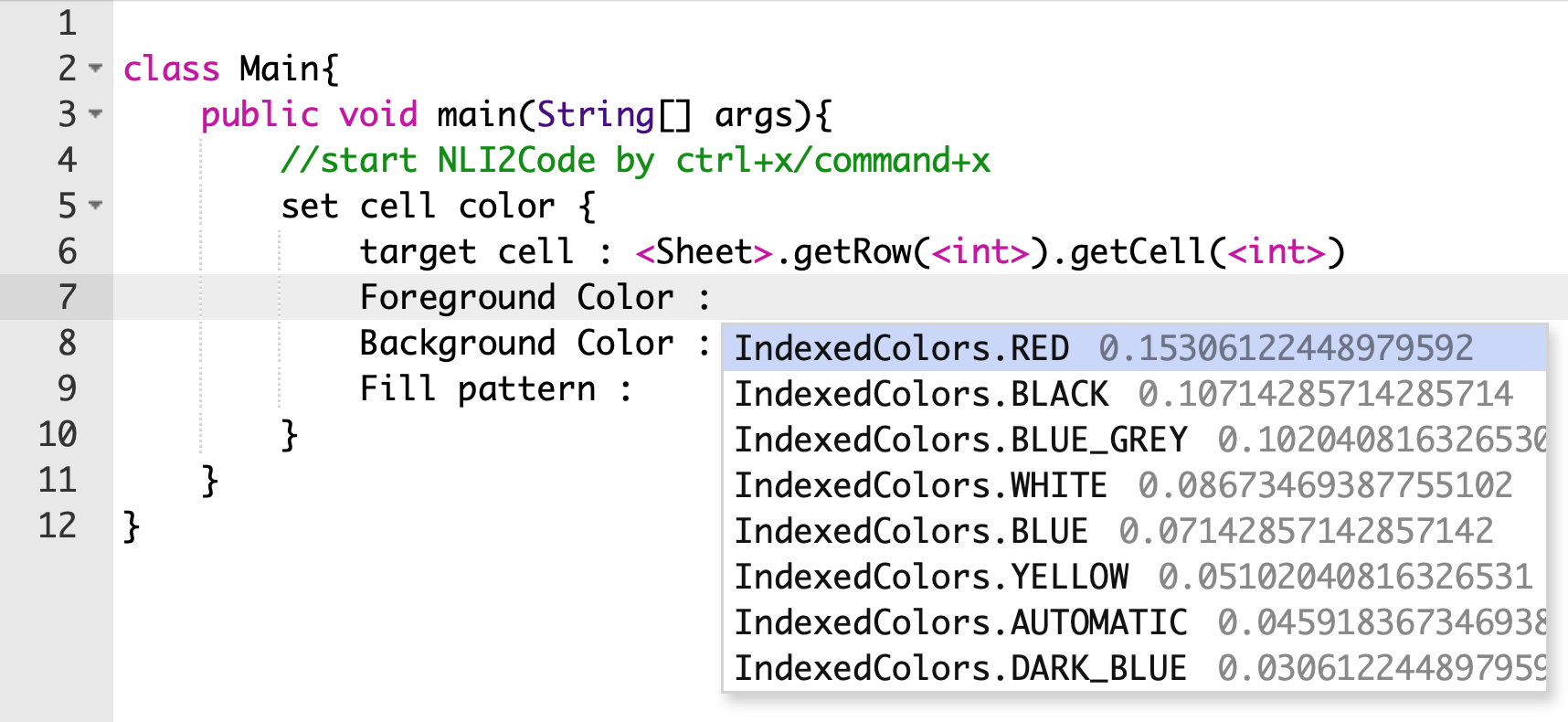}
\vspace{0.2cm}
\end{minipage}
}
\subfigure[The final complete code]{
\begin{minipage}{8cm}
\centering
\vspace{0.2cm}
\includegraphics[width=\textwidth]{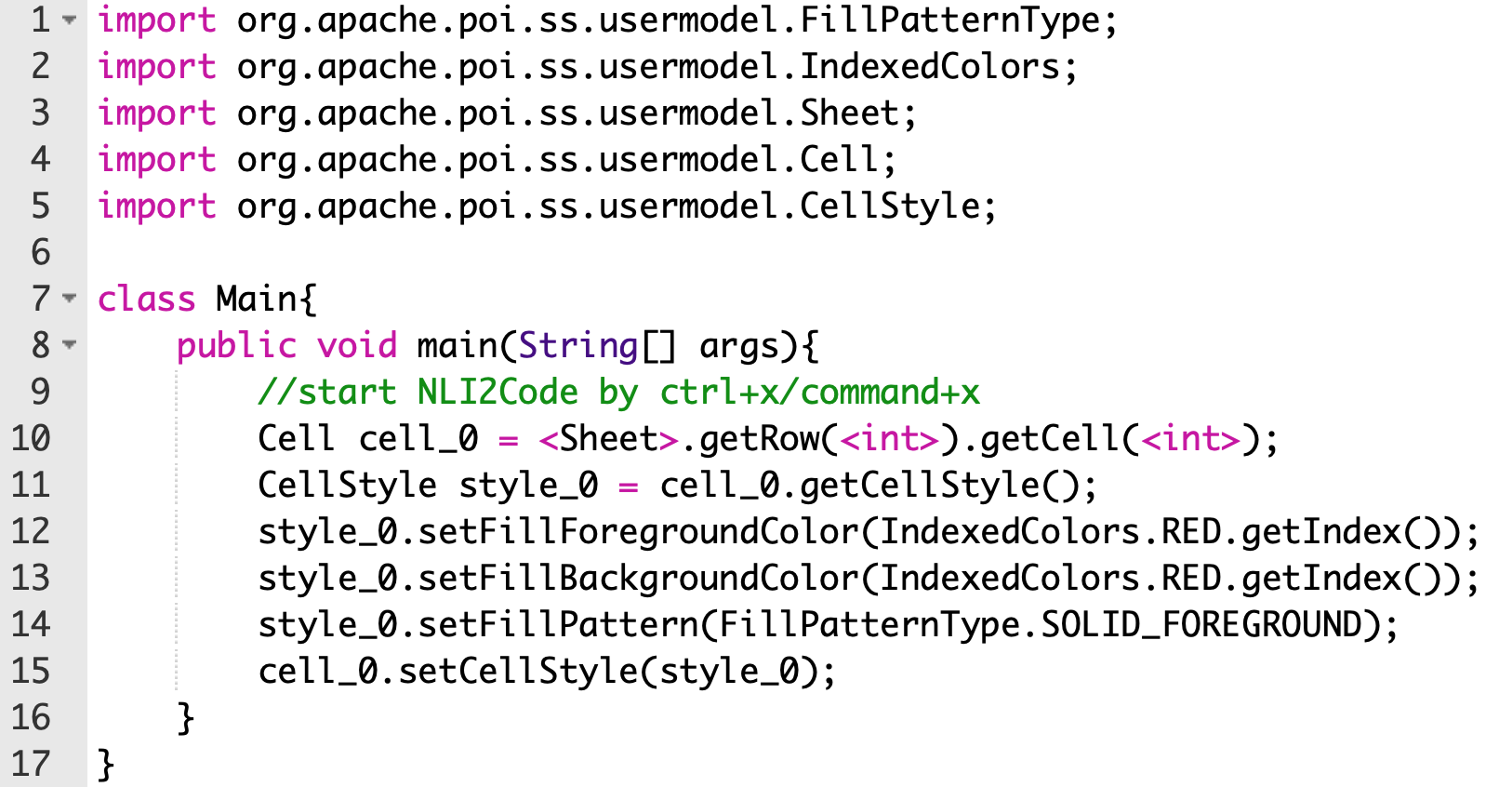}
\vspace{0.2cm}
\end{minipage}
}
\caption{User Interface of \textsc{CodA}}
\label{fig-ui}
\end{figure}

To ease the process of integrating API methods, we develop a tool that assists developers to integrate API usage patterns in an interactive way. 
Figure \ref{fig-ui} displays the user interface of our tool \textsc{CodA}. 
After triggering \textsc{CodA} with \texttt{ctrl-x}, users can select the desired task description, and our tool will hint users with the variables to configure. 
The hint for a variable contains a description for its meaning and a list of expressions to synthesize the variable, equipped with the occurring frequency in real-world usages. 
After users complete all the variables, a well-typed code snippet will be inserted at the current caret. 
The input of our tool is an API usage pattern to integrate, with some missing variables to complete. 
Given an API usage pattern, the tool first extracts its real-world usages from the code corpus. 
Then, the usages are analyzed to build the def-use relations between the receivers and parameters. The tool also records the expressions that declare our concerned variables. 
\textsc{CodA} annotates each variable with its description from Javadoc and recommends the frequent expressions to create the variable.
Besides, our tool also recommends a list of concrete API usage examples according to the current user configuration.

The contributions of this paper are:
\begin{itemize}
	\item A taxonomy of variable completion syntax for integrating API usage patterns. 
		We analyze 168 API usage patterns mined from five Java libraries, and conclude five syntax types for variable completion.
	\item An approach for comprehensive integration of API usage patterns, which guides users to interactively fill the
		missing variables.
	\item A code example recommendation study and a user study to evaluate the usefulness of our approach in real-world integration scenario.
\end{itemize}

%% file: sections/2_related.tex
\section{Related Work}
\label{section-related}
Since our approach is tightly associated with API usage pattern mining and API usage example recommendation, we first introduce some related works to help readers understand our work better.
\subsection{Mining API usage patterns}
An API usage pattern describes that in a certain usage scenario, some API methods are frequently called together and their usages follow some sequential rules\cite{zhong2009mapo}.
The general procedure of mining API usage patterns follows three steps.
First, a representative set of source code is collected as the code corpus.
Second, code is transformed into certain intermediate representation.
Common representations include call sequences\cite{zhang2018code}, execution traces\cite{yang2006perracotta}, syntax trees\cite{allamanis2014mining}, and graph structure\cite{nguyen2016t2api, DBLP:conf/sigsoft/NguyenNPAN09}.

The main application of the mined API usage patterns is to detect API misuse, which works by comparing a given API usage with the highest-ranked patterns and reporting the violations as misuse.
However, existing works seldom apply the mined patterns to generate well-typed code snippets.
Some works\cite{insynth, PLDI12:completion, pbe} discussed this problem by providing simple completion after mining API usage patterns.
SWIM\cite{raghothaman2016swim}, a program synthesizer tool for C\# completes a mined API usage pattern with simple rules(\textit{e.g.}, the default values for the basic types, directly call the constructor for the reference types).
NLI2Code\cite{shen2020api} combines type-directed search and user interaction to complete the missing parameters in Java API usage patterns.
Similar synthesis approach is also applied in some other works\cite{perelman2012type, gvero2013complete}.

Our conjecture is that users need more information than the default value, constructors, or synthesized expressions when filling an unfamiliar API usage pattern.
Recommendation of both candidate expressions and their usage in real-world code examples are valuable for developers.

\subsection{Recommending API usage examples}
Besides abstract code patterns mined from a large corpus, concrete code examples from the real world are also an important resource to learn APIs.
Recommending API usage examples can be viewed as a subfield of code search because it specifies the user query as an API element.

MUSE\cite{moreno2015can} accepts a given method and returns a list of its usage examples from the large corpus.
By applying static slicing and clone detection, MUSE can cluster similar examples and recommend different, representative usage of the interested method.
The tool further selects and ranks the examples with heuristics to improve the understandability and popularity of the recommendation result.
Some works\cite{autocomment, api-tutorial} notice that developers often need some descriptions accompanied with the returned code examples.
CROKAGE\cite{silva2019recommending} returns a comprehensive solution for a given programming task, containing both code examples and succinct explanations.
Furthermore, some works\cite{prompter, task2code, codehint} provide direct support for code search in IDEs.
CodeMend\cite{rong2016codemend} is a system that supports finding and integration of code, which leverages a neural embedding model to jointly model natural language and code.
CodeScoop\cite{head2018interactive} helps developers extract code fragments in their current projects for further code reuse.
Examplore\cite{glassman2018visualizing} is the state-of-art work for visualizing hundreds of examples for a given API element.
It predefines a synthetic skeleton for API usage, which includes seven API usage features including preconditions, return value check, exception handling, and so on.
Given a recommended API usage example, different features are marked with different colors to help developers quickly locate their desired part.
Similar to our work, Examplore\cite{glassman2018visualizing} can also update the order of recommended code snippets according to the current user configuration.
However, Examplore focuses on visualizing the API specification such as pre-conditions and exception handling.
Different from Examplore, our hypothesis is that API usage pattern mining tools can mine API specifications automatically, and we focus on how to configure a code example into a well-typed, executable snippet.

To sum up, recommending API usage examples is a hot topic in software engineering research.
Improving the reusability and understandability of examples is an important direction in this area.
In this paper, we further strengthen this by providing direct guidance during the API integration.

%% file: sections/3_analysis.tex
\section{Empirical Study on API Integration}
\label{section-analysis}
Before we dive into the details of our empirical study and approach, we first define two important concepts in this paper.
\begin{myDef}[API usage pattern]
An API usage pattern is a set of API methods that are frequently called in sequence.
There are some missing receivers or parameters of methods in an API usage pattern. 
After developers formulate the receivers and parameters, an API usage pattern becomes a complete code snippet.
\end{myDef}

\begin{myDef}[Hole]
The missing receivers and parameters of methods in the API usage pattern.
\end{myDef}

\subsection{Data collection}
To get insights into API integration, we collected 100 code elements and analyze their usage examples from Github.
Consider the ways of integration may vary among different APIs, the code elements are from five representative Java libraries, 
\textit{i.e.}, an html extraction library (\texttt{jsoup}), a source code parser (\texttt{eclipse-jdt}), a library manipulating Microsoft documents (\texttt{apache-poi}), 
a deep learning toolkit (\texttt{deeplearning4j}), and a graph database platform (\texttt{neo4j}).
In addition to being widely used, these five libraries cover different programming domains, from the front-end html parsing to the back-end database manipulation.
For each library, we use the library name as the query, search and download 500 client repositories from Github.
To guarantee the repository quality, each repository has at least two stars.
Repositories with fewer stars are removed, and we further remove forked repositories.
For each library, we parse the client repositories and count the occurrence of each code element from the library API.
We sort the code elements according to their occurrence times and remove trivial ones like \texttt{toString()}.
Finally, we collect the top 20 frequent code elements for each library.
For each code element, we collect 1,000 usage examples from the corpus.
A usage example for a code element is a method from the client repositories that invokes the code element.

\subsection{API usage pattern mining}
In this paper, we choose Structured Call Sequence (SCS) as the intermediate representation for code, which is also used in previous API pattern mining works\cite{raghothaman2016swim, zhang2018code}.
The syntax we used is the same as an existing work \textsc{ExampleCheck}, the detailed syntax rules can be found in the original paper.
There are two reasons we choose SCS as our code abstraction.
First, it has rich syntax to deal with the complex API usage model, including necessary control statements like \textit{if}, \textit{while}, and \textit{try} to represent preconditions and exception handling.
Second, there is a mature frequent subsequence mining algorithm PrefixSpan\cite{prefixspan} to efficiently mine the API usage patterns. 
Compared to other code abstraction forms, the sequence model is simple and scales to very large corpus.
One thing to note that, although this paper chooses SCS as code abstraction in implementation, our approach can be easily spread to other code abstractions, which only need to know the type of missing expressions in the mined patterns.

During mining, we choose the threshold as 5\%, which means a subsequence has to appear in at least 5\% of the files in the corpus.
Furthermore, we configured the BIDE tool to only return closed sequences, which means if the super sequence of a sequence is also frequent, this sequence won't be returned.
The minimum length of the returned sequence is 3.
The mining result only represent the subsequences are frequent, however, they do not necessarily represent a correct usage of a given API method.
Thus, we did a manual checking for each mined pattern and removed the following two types of results:
\begin{itemize}
\item \textbf{illegal syntax}. The mined pattern has illegal syntax, \textit{e.g.}, unclosed brackets
\item \textbf{unclear semantics}. The mined pattern is only a frequent combination of some unrelated code elements and does not have a clear functionality.
\end{itemize}

Finally, we mine 168 API usage patterns for the 100 chosen code elements.
We focus on an important problem during code integration: completing the missing variables of a given API usage pattern.
All the receivers and parameters of method calls in the mined SCS sequences need to be completed.
Totally, we collect 670 variables to complete.

\subsection{Completion type analysis}
\begin{figure}[t]
\centering
\begin{lstlisting}[language=Java]
class WriteExcelSheet {
public WriteExcelSheet(String path){
  try {
    f= new File(path);
    fis= new FileInputStream(f);
    wb= new XSSFWorkbook(fis);
  } catch (Exception e) {
    // lines of code ommited
  }
  public void writeData(){
    CellStyle <HOLE> = wb.createCellStyle();
    // lines of code ommited
  }
}
\end{lstlisting}
\caption{Example code to illustrate how holes are constructed}
\label{fig-codeexample}
\end{figure}

Given an API usage pattern and its usage examples, we implement a tool to automatically extract how holes in the pattern are filled.
For each hole in the pattern, the tool analyze the def-use relations in the usage examples and returns a list of expressions, corresponding to the real-world hole implementation.
We not only consider the internal data flow of a method, but the fields and the constructor functions as well.
Figure \ref{fig-codeexample} shows an example to illustrate how our tool works.
In this example, the \texttt{<HOLE>} with type \texttt{CellStyle} is filled from the \texttt{path} variable from the constructor function.
The following method call chain is called for the construction: \texttt{File(), FileInputStream(), XSSFWorkbook(), createCellStyle()}.
The first author initially inpected the returned expressions for 50 API usage patterns and already observed convergent completion types.
Then the author continued to inspect the returned expressions for all the 168 API usage patterns, since the list of completion types was converging.
This is a typical procedure in qualitative analysis\cite{berg2004qualitative}. 
Finally, we conclude 5 categories from the syntax perspective. 

\textbf{Enumerations} With enumerated constants, empty holes often indicate a software developer's need to choose from a limited set of options that best fit his or her needs.

\textbf{Method calls} Any method call with the correct return type can be a candidate expression for the hole. Furthermore, it could be a method call chain that returns the desired type. 
There are some noticeable ways of method calls to create the variables:
\begin{itemize}
	\item \textit{Design patterns}. Factory patterns and singleton patterns are the most common design patterns to create a new variable.
	\item \textit{Load persistent data}. Important data are often stored in persistant structure like databases or files. Load from such persistant storage is a common way for data creation.
	\item \textit{Getters}. A desired variable is often a private field of another data structure. A getter method is a common interface to visit such fields. 
\end{itemize}

\textbf{Constants} For the basic types, programming languages often provide their default value, such as 0 for \textit{int}, or "hello world" for \textit{string}, such types have infinite options, which can only be specified by users. For the reference types, the common constant is \textit{null}.

\textbf{Class instantiations} For the reference types, a variable could be instantiated with the corresponding constructors.

\textbf{Defined variables from the context} Unless a user decide to reuse an API usage pattern from scratch, the current programming context has already defined some variables. Such defined variables can be directly used to fill the holes.

Table \ref{tab-distribution} shows the distrubution of all the five types in our empirical study.
As we can see, the most frequent types are variables in the context and return value of method calls, cover 43\% and 37\% cases separately.
For completion of method call type, we observe that they often appear in the form of method call chain, such as \texttt{a().b().c()}.
Figure \ref{fig-methodchain} shows a code example to create a \texttt{Cell} type variable in \texttt{apache-poi}.
Such a long method call chain is difficult for users to understand and synthesize.

This problem arise the direction for our work: provide a comprehensive integration solution for users to complete the missing variables in API usage patterns.
We consider two ways to help users under such situations.
First, we observe that a common senario in API-centric code completion is that users have defined some variables, but don't know how to get a variable with the desired type.
Thus, we view the defined variables as the starting points to synthesize candidate expressions and divide the expressions according to their syntax type (among five types in this section). 
Second, we dynamically update a list of real-world API usage examples to users, to enhance their confidence that the recommended completion can work.

\begin{table}
    \centering
    \caption{Distrubution of the five types of completion types}
    \vspace{0.2cm}
    \begin{tabular}{l c c}
    \hline
	Type & \#Occurrences &Percentage(\%)\\
    \hline
	Enumerations & 25 & 3.7\\
	Method Calls & 248 & 37.0\\
	Constants &40 & 6.0\\
	Class Instatiation & 69 & 10.3\\
	Defined variables from the context & 288 & 43.0\\
    \hline
	\textbf{Total} & 670 & 100.0 \\
	\hline
    \end{tabular}
    \label{tab-distribution}
\end{table}

\begin{figure}[t]
\centering
\begin{lstlisting}[language=Java]
Workbook wb = new HSSFWorkbook();
Cell cell = wb.createSheet().createRow(0)
	.createCell(0);
\end{lstlisting}
\caption{A long method call chain to create a \texttt{Cell} type variable}
\label{fig-methodchain}
\end{figure}

%% file: sections/4_approach.tex
\section{Approach}
\label{section-approach}

\begin{figure*}[htb]
	\centering
	\includegraphics[width=0.85\textwidth]{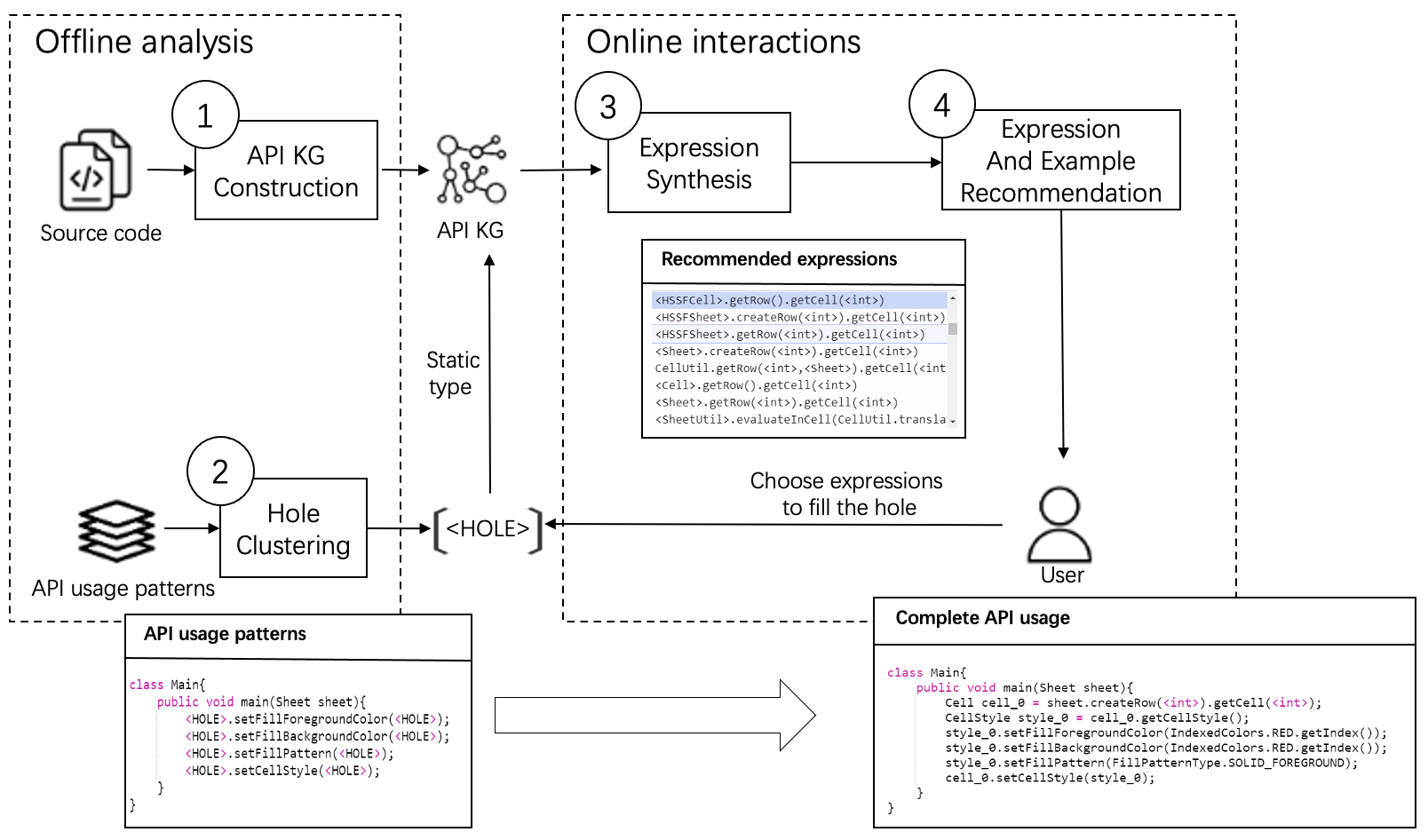}
	\caption{Framework of our approach}
	\label{fig-framework}
\end{figure*}

Figure \ref{fig-framework} shows the framework of our approach, which can be divided into four steps.
(1) API knowledge graph construction,
(2) hole clustering, 
(3) expression synthesis, and 
(4) expression and code example recommendation.
Given an API usage pattern, we first parse the API source code to construct a knowledge graph.
Since holes in a pattern may have data dependency, we further cluster the holes by analyzing the client code corpus.
The two phases are finished in the offline part.
In the online part, our goal is to guide users to complete the pattern into an executable code snippet.
By searching the knowledge graph, we synthesize well-typed expressions for each hole.
Also, a list of recommended code examples is shown for users to learn similar API usage.
After users select certain expression from our synthesis result, 
the code examples will be reranked to ensure the top examples are the most related to the current context.
The rest of this section will introduce the details in our approach.

\subsection{API knowledge graph construction}
Source code of a library contains abundant semantic knowledge, we focus on Java libraries in this paper:
\begin{itemize}
\item class abstracts real-world entity, method corresponds to operation on certain entity
\item mechanism like inheritance and \emph{java.lang.Iterable} interface models semantic relations between entities
\end{itemize}
Our approach organizes such semantic knowledge as the API knowledge graph (API KG).
Given the source code of a library, the construction of API KG is completely automatic and works in offline part of our system.

API KG is a directed graph $G=(V, E)$.
The node set $V$ contains five types of nodes, which correspond to five types of API elements.
All the nodes are components available for later expression synthesis.
We use ASTParser in org.eclipse.jdt.core\footnote{https://mvnrepository.com/artifact/org.eclipse.jdt/org.eclipse.jdt.core} 
to extract code elements and their semantic relations for a given API.

The edge set $E$ of API KG consists of six types of edges.
Extend and implement edges describe the inheritance property.
In Java, when we need an object of father class, we can provide a child class object.
The operation is called upcast and the reverse operation is called downcast.
We treat the implement edge in a similar way as the extend edge, which is also used to synthesize cast operations in the program.
The only difference is the target node of the implement edge is an interface, which can not be instantiated as an object.
For the iterable edge, \emph{java.lang.Iterable} is a built-in interface of Java.
If type \emph{A} implemented Iterable interface with type \emph{B} as parameter, variable with type \emph{A} can be iterated in a for-each loop.
Each element in the iteration has type \emph{B}.
An iterable edge starts from a class/interface node and ends with another class/interface node.
It means the source node implements the \emph{Iterable} interface with the target node as a parameter.
Details of the nodes and edges are shown in Table \ref{tab-node} and \ref{tab-edge}.

\begin{table}[htb]
\caption{Five Types of Node in API KG}
\label{tab-node}
\centering
  \begin{tabular}{l|p{2.5cm}|p{3cm}}
\hline
\textbf{Node Type} & \textbf{Description} & \textbf{Properties}\\
\hline
Class & a java class & name, comment\\
\hline
Interface & a java interface & name, super interface, comment\\
\hline
Method & a member method & name, parameter list, static or not, constructor or not, comment\\
\hline
Enum Class & an enum class & name, comment\\
\hline
Enum Constant & an enum constant& name\\
\hline
\end{tabular}
\end{table}

\begin{table}[htb]
\caption{Six Types of Edge in API KG}
\label{tab-edge}
\centering
  \begin{tabular}{l|p{1.5cm}|p{1.5cm}|p{3cm}}
\hline
\textbf{Edge Type} & \textbf{From} & \textbf{To} & \textbf{Description}\\
\hline
haveMethod & class/ interface/ enum class & method & a class/interface to its member methods\\
\hline
return & method & class/interface & a method to its return type\\
\hline
haveConstant & enum class & enum constant & an enum class to its member constants\\
\hline
implement & class/interface & interface & a class/interface to interfaces it implements\\
\hline
extend & class/interface & class/interface & inheritance of a class/interface to class/interface it extends\\
\hline
iterable & class/interface & class/interface & container class/interface to element class/interface\\
\hline
\end{tabular}
\end{table}

\subsection{Hole Clustering}
To integrate an API usage pattern into the local programming context, developers have to construct variables for all the receivers and parameters of the API methods.
It's common that variables in the API usage pattern have strong relationships with each other, 
which indicates that we can reduce the number of variables to be filled through investigating real-world usages.

Given an API usage pattern, we first view all receivers and parameters of methods are holes. 
Then we extract the actual receivers and parameters from real-world usages. 
By analyzing all actual parameters for a given hole, we can classify different holes into fixed or changeable. 
Only holes with a constant type such as string literals, numbers, and enums can be considered as fixed, 
so long as the frequency of the most frequent actual parameter exceeds a predefined threshold. 
Once a hole is considered as a fixed hole, the most frequent actual parameter will be filled in the hole. 
This process will repeat until all holes in the snippet skeleton are changeable.

Furthermore, we summarize relationships between the changeable holes and utilize them to reduce the number of changeable holes further. 
One of the simplest but important relationships is co-reference, 
which means that two different holes should refer to a single entity, or share the same expression as the parameter. 
A co-reference group is a set of holes in which the co-reference relationship exists between any two different holes in this group. 
We aim to find as many maximal co-reference groups as possible.
As a result, a maximal co-reference group can be viewed as a single hole and needs only a single expression as the parameter.

\begin{myDef}[Co-reference degree]
Co-reference degree between two different holes represents the frequency that the two holes are completed with identical expressions in the real-world usage.
Co-reference degree between two groups of holes represents the minimal value of co-reference degrees between holes in the two groups.
\end{myDef}

In the process of discovering maximal co-reference groups, a co-reference matrix was used as the key data structure.
Each line of this matrix refers to a co-reference group in the API usage pattern, and so do the columns. 
The i-th line and the i-th column represent the same group, and it makes the co-reference matrix a symmetric matrix. 
At first, each group has only a single hole, and any two groups in different lines contain different holes. 
The number resides in the i-th line and j-th column represents the co-reference degree between i-th group and j-th group. 
Based on the co-reference matrix, we discover maximal co-reference groups in an iterative, bottom-up way. 
In each iteration, our approach scans the matrix and find a grid in which the number exceeds a predefined threshold. 
Suppose the grid resides in the i-th line and j-th column, and it means that i-th group and j-th group have a co-reference relationship. 
Then we combine the two groups into a new group that contains all holes in the original two groups, and the co-reference matrix should be modified to reflect the combination. 
Specifically, the lines and columns corresponding to the combined two groups will be removed, and a new line and a new column will be inserted to the matrix, 
which represents the co-reference degrees between the new group and other remaining groups. 
According to the definition of co-reference degree, the new line can be computed as follows: 
the k-th entry in the new line is the minimum value between the k-th entries in the removed two lines. 
The computation of the new column is similar to the new line. 
After an iteration, the size of the co-reference matrix is reduced by one. 
The process will be repeated  until there is no value in the co-reference matrix exceeds the predefined threshold. 
At that time, no groups can be combined any more, so all the groups in the matrix are maximal.

\subsection{Expression synthesis}
Now we get a set of maximal co-reference groups for each API usage pattern, but there are no descriptions for these groups of holes. 
Without that, developers may have no idea what these groups represent, and thus they can not formulate appropriate expressions to complete the pattern. 
Therefore, it’s necessary to find a suitable description for each of the co-reference groups.

Our approach extracts descriptions from the Javadocs of the specified libraries. 
To be more specific, if a co-reference group corresponds to a parameter of a method invocation, we extract the parameter descriptions in the Javadocs 
(\textit{i.e.}, content after \texttt{@param} marks).
Otherwise, the group is a receiver of a method call, we use the class name as its description.

Even when there is a description for each group, developers sometimes still can not figure out how to formulate an appropriate expression for the group. 
For instance, to specify the red color in \texttt{apache-poi}, developers should use \texttt{IndexedColors.RED.getIndex()}.
Such an expression is difficult to be synthesized if the developer is not familiar with the library.
To address the problem, we proposed a method to synthesize expressions for each group, making use of both the source code and client code of the target library.

Recall we construct the API knowledge graph for a library in our offline approach.
We change the problem of synthesizing well-typed expressions to the problem of searching paths on the graph. 
Specifically, when a developer needs to formulate an expression for a given group, we firstly parse the local programming context to extract a list of local variables. 
Then we synthesize expressions according to the data type of the group in a top-down way. 
Algorithm \ref{algo1} shows the detailed procedure. 
For a given target type and maximal depth, we first add all local variables and static fields whose type equals to target type into the recommendation list. 
As for the non-static fields with target type, we then utilize Algorithm \ref{algo1} with \texttt{maxDepth-1} as a parameter to construct the objects which contain
target non-static fields, then take the target fields from the objects. 
For methods that return target type, we also utilize the algorithm with \texttt{maxDepth-1} as a parameter to get subexpressions for all parameters and callers of the methods, 
then combine the subexpressions and methods into a list of well-typed expressions. 
In this way, all expressions with target type can be synthesized, and the \texttt{maxDepth} parameter can be used to control the maximal complexity of synthesized expressions.
In our current implementation, we set \texttt{maxDepth} to 4. 
If the algorithm can not find any expression within \texttt{maxDepth} complexity, it will return only a placeholder expression of the target type.

\begin{algorithm}
\caption{Algorithm to synthesize a expression with a given type}
\begin{algorithmic} 
\Input
\Desc{localVariables, targetType, maxDepth}{}
\EndInput
\Output
\Desc{expressions}
\EndOutput
\State expressions = []
\If{maxDepth$\leq$ 0}
	\Return expressions
\EndIf
\State methods = getMethodsByReturnType(targetType)
\For{each method in methods}
	\State subTargetType = method.parameters + [method.caller]
	\State subExpressionsSequence = []
	\For{each subTargetType : subTargetTypes}
		\State subExpressions = synthesizeExptressions
		\State \quad(localVariables, subTargetType, maxDepth-1)
		\State subExpressionsSequence.append(subExpressions)
	\EndFor
	\State methodCalls = synthesizeMethodCallExpressions
	\State \quad(method, subExpressionsSequence)
	\State expressions.append(methodCalls)
\EndFor
\If{expressions.length == 0}
	\State expressions.append(placeHolder(targetType))
\EndIf
\end{algorithmic}
\label{algo1}
\end{algorithm}

\subsection{Expression recommendation and code example update}
After synthesizing a list of candidate expressions for each hole group, we need to recommend the expressions in a user-friendly way.
We divide the recommendation into two parts.

The first part is completion type selection, which is based on the conclusion of our empirical study.
Users will first select the desired completion type (\textit{i.e.}, enumerations, constants, class instantiations, method calls, and defined variables).
For enumerations, we let the user select among a fixed set of possible enumeration constants.
For example, for the missing parameter of method \texttt{setFillPattern}, our API KG shows that there are 19 patterns defined in \texttt{apache-poi}.
Thus, a drop-down selection box is displayed for users to choose the desired pattern.
For constants such as a user-specified string or magic number, an input box is displayed for the user to type the content.
For class instantiations, we combine the constructors and some classic design patterns for class instantiation under the category.
Users simply choose which one they need in a selection operation, just like working with enumerations.
For method calls and defined variables, we treat them as the same because they all start from defined variables and try to synthesize the desired type.

The second part is expression ranking. Among expressions of the same completion type, the following two metrics are used to rank them.

\textit{Completeness} For API-centric code completion, a common scenario is that 
the programmer knows what type of object he needs, but does not know how to get the object with variables already defined\cite{mandelin2005jungloid}.
Thus, according to the number of undefined variables in an expression, we rank more complete expressions higher.

\textit{Popularity} Since the completeness metric often brings two expressions to a tie, we utilize the client code of the library to view the popularity of the synthesized expressions. 
The score of an expression is computed as the product of the scores of its components.
The basic elements of an expression are methods, fields, constructors, local variables, and placeholders. 
To get a variable of a given type, developers can utilize a constructor, a method that returns a variable with the given type, or a field of that type. 
We analyze the client code and estimates the probabilities of the three ways to create a variable. 
This also implies that the scores of all methods, constructors, and fields with the same return type will add up to one. 
As for local variables, we do not penalize their existence and set their scores to one.
In this fashion, the expressions are ranked according to their occurring probabilities.

%% file: sections/5_evaluation.tex
\section{Evaluation}
\label{section-evaluation}

We focus on the following two research questions:
\begin{itemize}
	\item \textit{RQ1.} Given a specified code example among the code corpus, can \textsc{CodA} recommend it effectively with user interaction?
	\item \textit{RQ2.} Can \textsc{CodA} save time and help developers comprehend the API usage better during integrating real-world API usages?
\end{itemize}

We conducted two evaluations to answer the questions. 
For the first research question, we randomly pick a program from the corpus for API usage pattern mining as our goal example.
We suppose the user wants to do the same thing as the goal program, and simulate the user's completion process.
The ranking of the goal example is recorded for analysis.
This evaluation gives a feel for the accuracy of the algorithm.
The second question is evaluated in a user study. 
We prepare five curated API usage examples from Stack Overflow and ask participants to use \textsc{CodA} to integrate them.
The user study shows how our tool works in real-world development and detects how it can promote the efficiency of API integration.

\subsection{Datasets}
We use 3 Java libraries in our evaluation, which include:
Apache POI, a library manipulating Microsoft files like Word and PowerPoint. 
Joda-time, a library process time, which is later integrated into the JDK.
JFreeChart, a library to generate different kinds of charts.

Table \ref{tab-datasets} shows how many class files and non-blank lines of code each project contains.
We also reported the time cost to generate our API knowledge graphs.

\begin{table}
    \centering
    \caption{Project and the corresponding API KG statistics}
    \vspace{0.2cm}
    \begin{tabular}{l l l l l}
    \hline
    Project & \#Classes & \#Methods & API KG time(s)\\
    \hline
	Apache POI & 3430 & 32274 & 268\\
    Joda time & 449 & 9785 & 14\\
    JfreeChart & 908 & 11017 & 60\\
    \hline
    \end{tabular}
    \label{tab-datasets}
\end{table}

\subsection{RQ1. Code example recommendation study}

\subsubsection{Methodology}
To evaluate our approach, we feed our tool with API usage examples from the official tutorial of each library. 
We choose examples from the official tutorials as input because they are widely adapted to the client code and make the process of mining API usage patterns easy.
We first construct the adaptation corpus for each example by detecting the same API call sequence in Github.
We apply \textit{Baker}\cite{liveapi}, a tool to automatically complete the fully-qualified names of the API methods.
For each example, we construct a corpus containing 100 adaptation examples.
We further select a specific adaptation for each example as our goal.
We feed the corpus to our tool and check whether our tool can promote the ranking of the goal example after several rounds of interaction.
To be specific, the goal example is randomly ranked in the 100 adaptation examples at the beginning.
For each hole of the API usage pattern, the goal example has its way to construct the missing variable.
Iteratively, we solve each hole in the pattern by providing how it is filled in the goal example.
This is a simulation of how developers use \textsc{CodA} to view their desired code examples.
We also first tell \textsc{CodA} the syntax type to fill one hole, then we automatically select the best completion way from \textsc{CodA}'s synthesized expression list.
The iteration process repeats until all the holes are filled.

\subsubsection{Result}
Table \ref{tab-rq1} shows the result for the three Java libraries.
The \#patterns column shows that the number of code examples in the tutorial for each library.
In total, we collected 93 code examples.
Each code example has 100 adapted examples, we randomly select 10 out of them as the set of goal examples.
That is to say, our artificial corpus evaluated \textsc{CodA} on integrating 930 API usage examples.

The Rank@Hole1, Rank@Hole2, and Rank@Hole3 columns show how \textsc{CodA} promotes the ranking of the goal example.
(\textit{e.g.}, Rank@Hole1 is the ranking of the goal example after the user fills the first hole, Rank@Hole2 is the ranking after two holes are filled.)
Finally, the Final Rank column represents the ranking after filling all the holes.
As we can see, each iteration significantly promotes the ranking.
After the user correctly fills all the holes, the desired code examples are displayed in top-5 on average.
For JfreeChart, the final ranking is 2.7, which means the users can expect to see the goal example in the top-3 recommendation of \textsc{CodA}.
Consider that there are 100 candidate code examples in the code corpus, \textsc{CodA}'s performance of recommending related code examples is effective.

\begin{table}
    \centering
    \caption{How the ranking of the goal examples are promoted by \textsc{CodA}}
    \vspace{0.2cm}
    \begin{tabular}{l l l l l l}
    \hline
    \multirow{2}{*}{Library} & \multirow{2}{*}{\#Patterns} & Rank & Rank & Rank & Final\\
    & & @Hole1 & @Hole2 & @Hole3 & Rank\\
    \hline
    Apache POI & 46 & 48.2 & 31.4 & 12.8 & 3.9\\
    Joda time & 28 & 43.3 & 27.8 & 14.6 & 4.3\\
    JFreeChart & 19 & 46.1 & 25.5 & 10.8 & 2.7\\
    \hline
    \end{tabular}
    \label{tab-rq1}
\end{table}

\subsubsection{Discussion}
The goal of our first research questions is to determine whether \textsc{CodA} can work on real-world Java libraries. 
The results suggest that the problem is tractable: our search and synthesis process on API KG achieves a modest speed and can effectively bring the goal code example to users. 
On average, the recommendation for a hole can be solved in less than 800 milliseconds, which is acceptable for an autocomplete tool. 
Note that we didn’t include the time to build the API KG since the model can be constructed in the offline process. 

\subsection{RQ2. User study}
The purpose of this study is to test the usefulness of \textsc{CodA} in real-world programming with human interactions.
We conducted a user study with eight Java programmers to evaluate whether participants could grasp a more comprehensive view of API usage using \textsc{CodA}, 
in comparison to a realistic baseline of \textit{Codota}, which is a widely-used deep learning based code completion plugin. 
\textit{Codota} can also recommend relate code examples if users give a specific API method name, which is commonly used in real-world programming workflows. 

\subsubsection{Participants}
We invited eight participants to solve real-world programming tasks using \textsc{CodA}. 
All the participants were familiar with the Java syntax.
Four of them come from the industry, with at least four years of experience in industrial software development.
The other four participants are graduate students majoring in software engineering.

\subsubsection{Tasks}
We designed a set of API integration tasks to assess how much knowledge about API usage participants could extract from our tool. 
Prototypes of the tasks were selected from Stack Overflow posts, which are displayed in Table \ref{tab-questions}.
In fact, adapting existing code snippets from Stack Overflow is a common scenario in software development.
Related works\cite{wu2019developers, treude2017understanding} show that the utilization of curated code examples from online forums is low due to multiple problems like incompleteness, incomprehensive.

Formally, each task in our user study consists of four components:
\begin{itemize}
	\item A code snippet to be integrated. Participants need to integrate the snippet into a given programming context.
		The code snippets are carefully selected to ensure that they display the correct API usage.
	\item A textual description of the task. We extract the title and surrounding sentences of the code snippet from Stack Overflow.
		We further concretize the description to give participants necessary information to integrate the snippet 
		(\textit{e.g.}, for Q4, we specify the position of the cell and the color to set).
	\item A programming context. One solution is that we can leave an empty context for participants to adapt the code snippets.
        However, to estimate a real-world programming environment, we collect the adaptations of the code snippet from Github.
        All the five snippets are internal calls in a single method, and we use the most frequent method signature from the adaptations as the context in the study.
    \item A testing program. For each task, we manually compose a testing program to validate the correctness of integration.
		For example, after specifying the position of the cell and the color to set, our testing program reads the color information of the specified cell and check whether 
		it is corrected configured. The testing process is completely automatic.
\end{itemize}

During the selection of the tasks, we follow the following two principles. 
First, the integration should focus on API usage instead of other configuration.
For example, APIs related to database connection requires users to correctly configure a database before composing a piece of correct code, which is not the target of this user study.
Second, we avoid APIs that participants are already familiar with before the study.
Thus, we select API usages from third-party libraries instead of JDK methods.

\begin{figure*}
    \centering
    \includegraphics[width=0.9\textwidth]{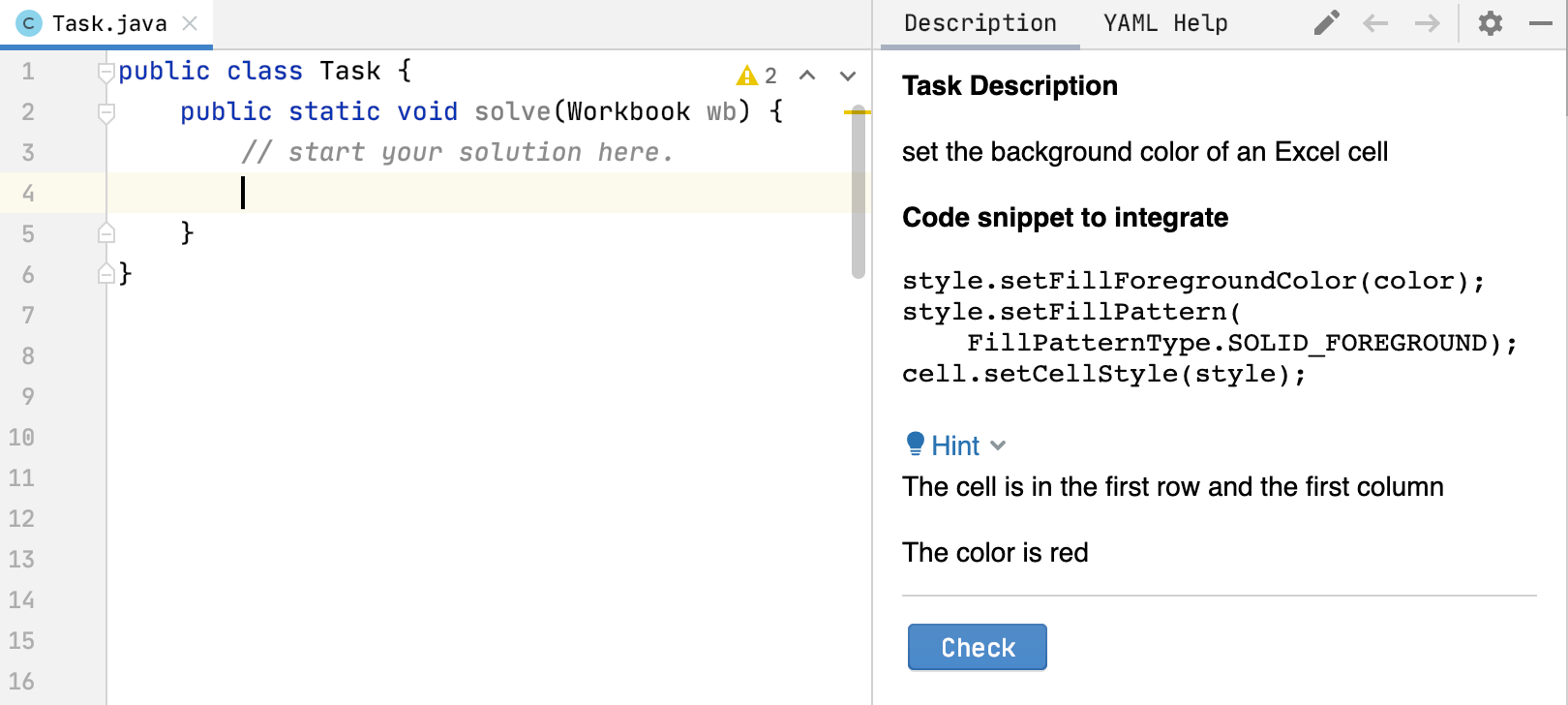}
    \caption{User interface of our user study. In this task, user are provided with a method signature, containing a parameter with the type \texttt{Workbook}.
    After user integrate the code snippets on the right side, a click on "Check" will return the verdict result.}
    \label{fig-studyui}
\end{figure*}

Finally, we use the EduTools of IntelliJ IDEA to implement the environment of our user study.
All the participants are given a static method \texttt{foo} without a method body, and they are asked to complete the method by integrating a set of APIs from the task.
Figure \ref{fig-studyui} shows the user interface of our user study.
A task is considered to be finished if the user can pass the testing program in a limited time (15 minutes), otherwise, the task fails.

\begin{table}
    \centering
    \caption{Study questions from Stack Overflow}
    \vspace{0.2cm}
    \begin{tabular}{l}
    \hline
    Task Descriptions\\
    \hline
    Q1. How to create hyperlink within an excel cell\\
    Q2. How to set time zone with Joda-time\\
	Q3. How to create an excel drop down list\\
    Q4. How to set the background color of an Excel cell\\
    Q5. How to display value in pie chart\\
    \hline
    \end{tabular}
    \label{tab-questions}
\end{table}

\subsubsection{Methodology}
We design the user study as a controlled study.
Two industrial programmers and two students form the experimental group, use \textsc{CodA} to integrate code snippets, 
while the control group consisting of the rest four participants don't use our tool.
Instead, the control group's coding environment is equipped with \textit{Codota}, a widely-used deep learning based code completion plugin.
\textit{Codota} can also recommend a list of code examples if the user types in the interested API method name.
We allowed all participants to visit online resources such as Q\&A forums and search engines when solving tasks.

During the user study, we record information of four aspects for later analysis, which are:
(1). the rank of the correct expression in our recommendation list, 
(2). the times of user interaction to finish a task, 
(3). the total time a participant used to finish a task, and 
(4). the response time of our tool to recommend expressions and examples.
The metric we use to evaluate the recommendation accuracy is MRR (Mean Reciprocal Rank), 
which is a typical metric used in IR (information retrieval) works.
In our scenario, completing a hole is an IR task.
For a set of holes $H$, the values of MRR is calculated as follows:
\begin{equation}
\begin{aligned}
score(hole) =
\begin{cases}
\frac{1}{rank(answer(hole))}& \text{answer is synthesized}\\
0& \text{else}
\end{cases}
\end{aligned}
\end{equation}

\begin{equation}
MRR(H) = \frac{\sum_{h\in H}{score(h)}}{|H|}
\end{equation}

For the user interaction times, we consider the following five actions as user interaction:
(1). fill in a method parameter, 
(2). fill in the receiver of a method, 
(3). fill in a method name, 
(4). fill in a class name, 
(5). instantiate an object.
Response time refers to the time spent by our tool for synthesizing and ranking expressions, 
starts from when the developer triggers \textsc{CodA} and finishes when the recommended expressions are returned to the user. 

\subsubsection{Result}

\begin{table}
    \centering
    \caption{Result for the experimental group}
    \vspace{0.2cm}
    \begin{tabular}{l l l l l l}
    \hline
    Task ID & \#Holes & \#Interactions & MRR & Total time(s) & Response time(s)\\
    \hline
    Q1 & 9 & 4 & 0.75 & 377 & 0.67\\
    Q2 & 10 & 3 & 0.83 & 456 & 0.86\\
    Q3 & 9 & 9 & 0.25 & 768 & 0.29\\
    Q4 & 10 & 4 & 0.75 & 143 & 0.55\\
    Q5 & 13 & 8 & 0.56 & 123 & 0.43\\
    \hline
    \end{tabular}
    \label{tab-studyresult}
\end{table}

Table \ref{tab-studyresult} shows the result for the experimental group in our user study. 

The \#Ele column represents the number of code elements that need to be completed to generate a complete code from the code skeleton without using \textsc{CodA}.
The \#Interact column represents the number of interactions required to generate a complete code from the code skeleton using \textsc{CodA}. 
It can be seen that using this tool can reduce the number of interactions by 50\%-67\% in most cases, and can greatly improve the efficiency of software developers.

From the MRR column, we can see that our tool has a high precision of recommending related expressions.
However, we notice that Q3 has a significantly poor result on the MRR metric, that is because the task requires initializing an array, which is currently out of the ability for our expression synthesizer.

\begin{figure}
    \centering
    \includegraphics[width=0.4\textwidth]{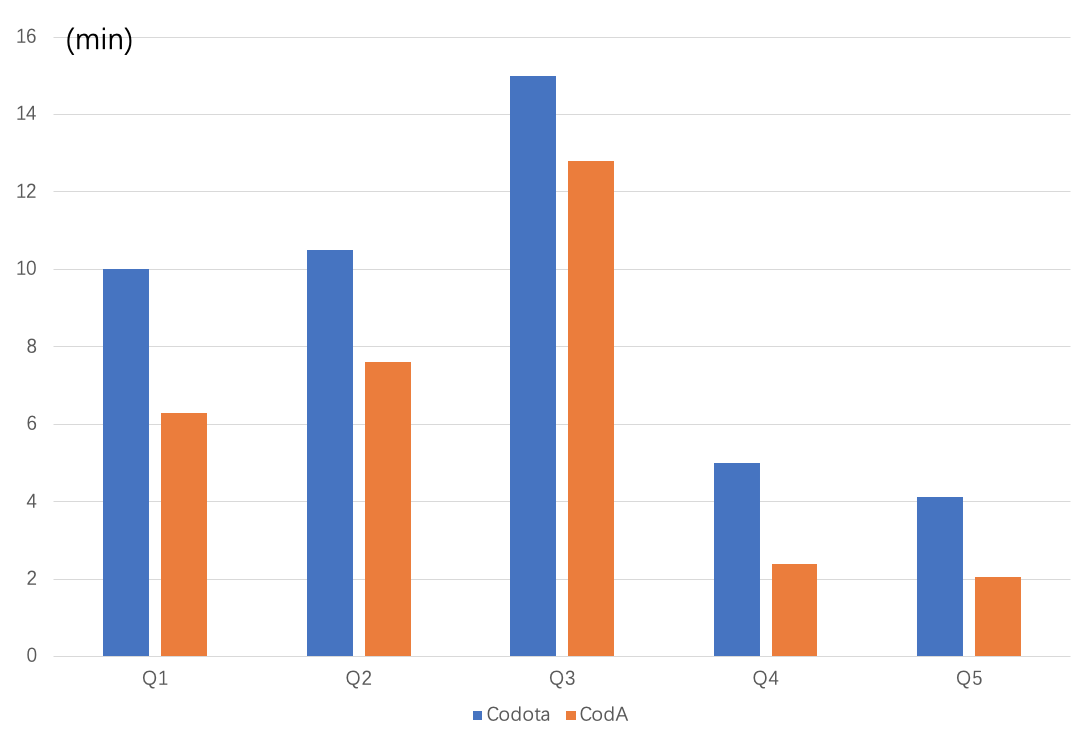}
    \caption{Comparison of average time spent by the two groups}
    \label{fig-timecompare}
\end{figure}

Figure \ref{fig-timecompare} compares the average time (minutes) used between the two groups.
On average, the control group spent 535.2 seconds for each task, which is significantly larger than the number for the experimental group (373.4 seconds). 
For Q3, two of the participants using Codota fails to pass the testing program, the other two almost spent all the limited 15 minutes.
We address the main advantage of \textsc{CodA} is that it can automatically update the related code examples, in contrast to Codota, which asks the user to type in a specific API method name and return a fixed list of API usage examples.

\subsection{Threats to validity}
\textbf{Internal validity} Due to the high cost of user study, we select only five tasks from Stack Overflow in our evaluation.
The performance of \textsc{CodA} may vary in other scenarios or real-world development.
To issue the problem, we automatically simulate the user's behavior on integrating 930 real-world API usage examples.
However, we cannot cover all the aspects of \textsc{CodA} in the automatic evaluation, instead, we only analyze how the recommended code examples change.

\textbf{External validity} We select three Java libraries from different domains to evaluate the usefulness of \textsc{CodA}.
The number of libraries is relatively small and may not be representative for all Java APIs.
The scale of a library and its client code may have effect on \textsc{CodA}'s performance.
First, if a library contains too many code elements, not only the time for offline API KG construction will be longer, the speed of online expression synthesis and recommendation may be extended as well.
In our datasets, \texttt{apache-poi} is the largest library, which consists of more than 3,000 classes and 30,000 methods.
For library of such size, \textsc{CodA}'s speed is acceptable (268 seconds for offline API KG construction, 800 milliseconds for each online recommendation).
Second, the scale of the client code is also an important factor.
Notice that not all open-source libraries are actively reused in platforms like GitHub.
If \textsc{CodA} is constructed for a relatively unpopular library, it may not find enough usage examples for recommendation.
\textsc{CodA} is designed to help developers quickly locate desired code examples by synthesizing missing variables interactively.
However, if the number of code examples is small at the beginning, the benefits of \textsc{CodA} will not be that significant.

%% file: sections/6_conclusion.tex
\section{Conclusion}
\label{section-discussion}
In this paper, we promote an approach to integrate a set of API methods by synthesizing their receivers and parameters.
Furthermore, our tool \textsc{CodA} can update a list of recommended real-world code examples according to the user's current context.
Our work is complementary to existing works on mining API usage patterns or synthesizing incomplete API usage patterns.
Instead of view the completion of method receivers and parameters as a straightforward code completion task, we address that users' comprehension of the APIs is important during the process.
Our tool gives a more comprehensive solution that existing AI-based completion tools because it classifies the recommended expressions according to their syntax, accompanied with a dynamically-updated code example list.

%% file: sections/7_acknowledge.tex
\section*{Acknowledgement}
This work is supported in part by the General Program of National Natural Science Foundation of China (61972006).